%% file: main.tex
\documentclass{svproc}
\usepackage[utf8]{inputenc}
\usepackage{subfiles} 
\usepackage{mathrsfs,amsmath}
\usepackage{algpseudocode}
\usepackage{algorithm}
\usepackage{multicol}
\usepackage{cite}

\date{April 2022}

\usepackage{url}

\usepackage{graphicx}
\usepackage{caption}
\usepackage{amsmath}

\usepackage{subcaption}
\usepackage[english]{babel}

\graphicspath{ {./figures/} }
\begin{document}
\mainmatter              
\title{At-the-edge Data Processing for Low Latency High Throughput Machine Learning Algorithms}
\titlerunning{At-the-edge Data Processing for Machine Learning}  
%
\author{Jack Hirschman\inst{1,2,*} 
\and Andrei Kamalov\inst{2} 
\and Razib Obaid\inst{2} 
\and Finn H. O'Shea\inst{2} 
\and Ryan N Coffee\inst{2}
}
\authorrunning{Jack Hirschman et al.} 
%
\tocauthor{Jack Hirschman, Andrei Kamalov, Razib Obaid, Finn H. O'Shea, and Ryan N Coffee}
\institute{Stanford University, Stanford, CA 94305, USA,\\
\and
SLAC National Accelerator Laboratory, Menlo Park, CA 94025, USA\\
\email{*jhirschm@stanford.edu}}

\maketitle              

\begin{abstract}
High throughput and low latency data processing is essential for systems requiring live decision making, control, and machine learning-optimized data reduction.
We focus on two distinct use cases for in-flight streaming data processing for a) X-ray pulse reconstruction at SLAC’s LCLS-II Free-Electron Laser and b) control diagnostics at the DIII-D tokamak fusion reactor.
Both cases exemplify high throughput and low latency control feedback and motivate our focus on machine learning at the edge where data processing and machine learning algorithms can be implemented in field programmable gate array based hardware immediately after the diagnostic sensors.
We present our recent work on a data preprocessing chain which requires fast featurization for information encoding.
We discuss several options for such algorithms with the primary focus on our discrete cosine and sine transform-based approach adapted for streaming data.
These algorithms are primarily aimed at implementation in field programmable gate arrays, favoring linear algebra operations that are also aligned with the recent advances in inference accelerators for the computational edge.
\end{abstract}
\section{Introduction}
\subfile{sections/introduction}


\section{Algorithms}
\subfile{sections/algorithms}


\section{Results}
\subfile{sections/results}
\section{Conclusions}
\subfile{sections/conclusions}

\section{Acknowledgements}
This research is supported by the Department of Energy, Office of Science, Office of Basic Energy Sciences for funding the development of the CookieBox detector array itself under Grant Number FWP 100498 ``Enabling long wavelength Streaking for Attosecond X-ray Science'' and for funding Field Work Proposal 100643 ``Actionable Information from Sensor to Data Center'' for the development of the associated algorithmic methods, EdgeML computing hardware, and personnel. We also acknowledge funding for the computational method development for tokamak diagnostics by the Office of Fusion Energy Science under Field Work Proposal 100636 ``Machine Learning for Real-time Fusion Plasma Behavior Prediction and Manipulation.'' Also, this research is funded by the Department of Defense National Defense Science and Engineering Graduate Fellowship. 

%
%

\subfile{sections/bib}
\end{document}

%% file: sections/introduction.tex
Data preprocessing for low-latency, high-throughput machine learning algorithms often requires efficient peak finding.
There are many applications that could utilize this particular preprocessing. 
We focus on two such applications in this manuscript: a) analyzing electron time-of-flight spectra for X-ray pulse reconstruction at SLAC's LCLS-II X-ray Free-Electron Laser (xFEL) in a detector called the CookieBox \cite{cookiebox} and b) analyzing Alfv\'en eigenmode frequency information in electron cyclotron emission (ECE) at the DIII-D tokamak fusion reactor \cite{toka}.
For the xFEL case, the arriving signals are roughly 0.5 ns in width, sampled at 6 GSa/s, and arrive over the course of 500 ns. 
The signals within that 500 ns window are not piled up, and, in general, are sparsely distributed. 
The time stamps of the peaks represent arrival times of electrons at the detector and can be used to compute the electron energies. 
So, a precise estimate for the arrival time of the electron will directly affect the energy calculation and thus the resulting machine learning-based X-ray pulse reconstruction.

For the fusion ECE case, a time-domain signal from a multi-channel radio-frequency pickup array is first transformed into a spectrogram where well defined frequencies appear as very narrow features in the 10--200 kHz range. 
These sharp frequencies are sparsely distributed throughout the spectral range and so an approach identical to the one used in the Cookiebox case is used to locate--and sub-sample--these sparse features.
This procedure significantly reduces the number of coefficients required to represent the spectrum.
Therefore, in both use cases the problem boils down to a peak-locating algorithm that can be applied early in the signal processing pipeline, before aggregation into a multi-channel image-like representation.

There are several approaches to general peak-finding from the most basic constant fraction discrimination \cite{peakFind:thesis, peakFind:old,peakFind:simple} to more sophisticated iterative curve fitting methods \cite{iterative, iterPaper, iterPaperGood}.
However, these iterative algorithms tend to require higher precision, significant run-time for convergence, and are generally less friendly for hardware targeted low-latency data processing implementations which significantly benefit from deterministic latency. 
Taking inspiration from the very efficient historical analog electronics like EG\&G Time-to-Digital Converter (TDC)\cite{egg}, we aim to use the similar approach to enhancing the resolution of peak finding for the general case, not only for the time domain signals.
Since we have the freedom to use our representation mappings to move the desired signal into the known coefficients, we are also free to apply concepts like Wiener filtering \cite{wiener} to these deeper representations, but doing so in the same algorithmic operation that results in the location itself.
This composite algorithm transforms a relatively high-dimensional, sparse data stream into a lower-dimensional, information-rich representation for downstream inference models.
Here we take the approach of using data streaming methods that leverage the capabilities of our hardware and which require a simpler peak finding method.
We focus on algorithms for computing derivatives that are employed for finding zero crossings and thus the corresponding extrema in the original data.

We further show that our efficiency in information extraction also allows us to reduce the bit-depth needed to represent these coefficients.
The reduced bit-depths conserve the often limited resources found in the FPGA hardware available for processing beyond this algorithm. 
The benefit of enforcing parsimony in the algorithm and the representation also yields very low latency when coupled with non-iterative and deterministic peak finding procedures.
In this paper we compare two algorithms, a convolution method and a discrete cosine and sine transform method (DCSTM), for identifying extrema in data streaming directly from detectors.
The algorithms are designed for deployment on hardware, but are first modeled here in software for development.
In both cases, we identify peaks by looking for zeros in the first derivative of the signals.

%% file: sections/algorithms.tex
This section discusses the algorithms developed for our data processing chain. 
We discuss two algorithms for calculating the derivative, the convolution method and the discrete cosine and sine transform method (DCSTM), as well as the underlying streaming discrete cosine transform method (SDCTM) used for calculating the discrete cosine transform (DCT), the discrete sine transform (DST), the inverse discrete cosine transform (IDCT), and the inverse discrete sine transform (IDST).
We also discuss approaches for finding the peaks of the original data via zero crossings in the output of a derivative.
The design of our algorithms are all focused on implementation in hardware.
We prioritize designs that will minimize latency and strive for efficient resource usage in an FPGA.

\subsection{Convolution Method}
The convolution method uses a stationary kernel function and streams data past this kernel while the data interacts with the kernel.
The choice of the kernel will change the operation at hand.
We want to choose a kernel that can take the derivative of our input signal. 
We arrive at the desired kernel by two approaches.
For the first, we use the 1-D derivative property of convolution \cite{Heckbert, boxlets}.
\begin{equation}
\label{eq:derConv}
    \frac{d}{dx}\left(f\ast g\right) = \frac{d}{dx}\left(f\right)\ast g = f \ast \frac{d}{dx}\left(g\right)
\end{equation}
Equation \ref{eq:derConv} shows we can take convolution of the derivative of the kernel $g$ with our original signal, and this result would be equivalent to taking the convolution of the derivative of our signal with the kernel $g$. 
Now if we consider a Gaussian kernel, $g$, then when we convolve it with the derivative of our input signal, we get a smoothed version of our input signal \cite{Talbi}.
Therefore, we can actually compute the derivative of this Gaussian kernel and convolve it with the original signal in order to get the smoothed version of the derivative of our signal.

For this, start by choosing a Gaussian kernel 
\begin{equation}
    g  = \frac{1}{ \sqrt{2\pi}}e^{{\frac{-t^2}{2\sigma^2}}}
\end{equation}
and then take the derivative to get
\begin{equation}
    \label{eq:conv}
    h = \frac{-t}{\sigma^2 \sqrt{2\pi}}e^{{\frac{-t^2}{2\sigma^2}}},
\end{equation}
where $\sigma$ is the kernel width, and finally convolve the kernel with the input signal. 

The second line of reasoning to arrive at this kernel shape is to consider looking at the noiseless power spectrum of the data of interest and then multiply it by the fourier transform of a step function which represents a derivative \cite{book1, website}. 
The resulting multiplication is then inverse fourier transformed to arrive at the analytic derivative of the Gaussian.
Section \ref{sec:convResults} discusses this second line of reasoning in the context of choosing the Gaussian derivative kernel width.

\begin{figure}[h!]
    \centering
    \includegraphics[width=0.55\textwidth]{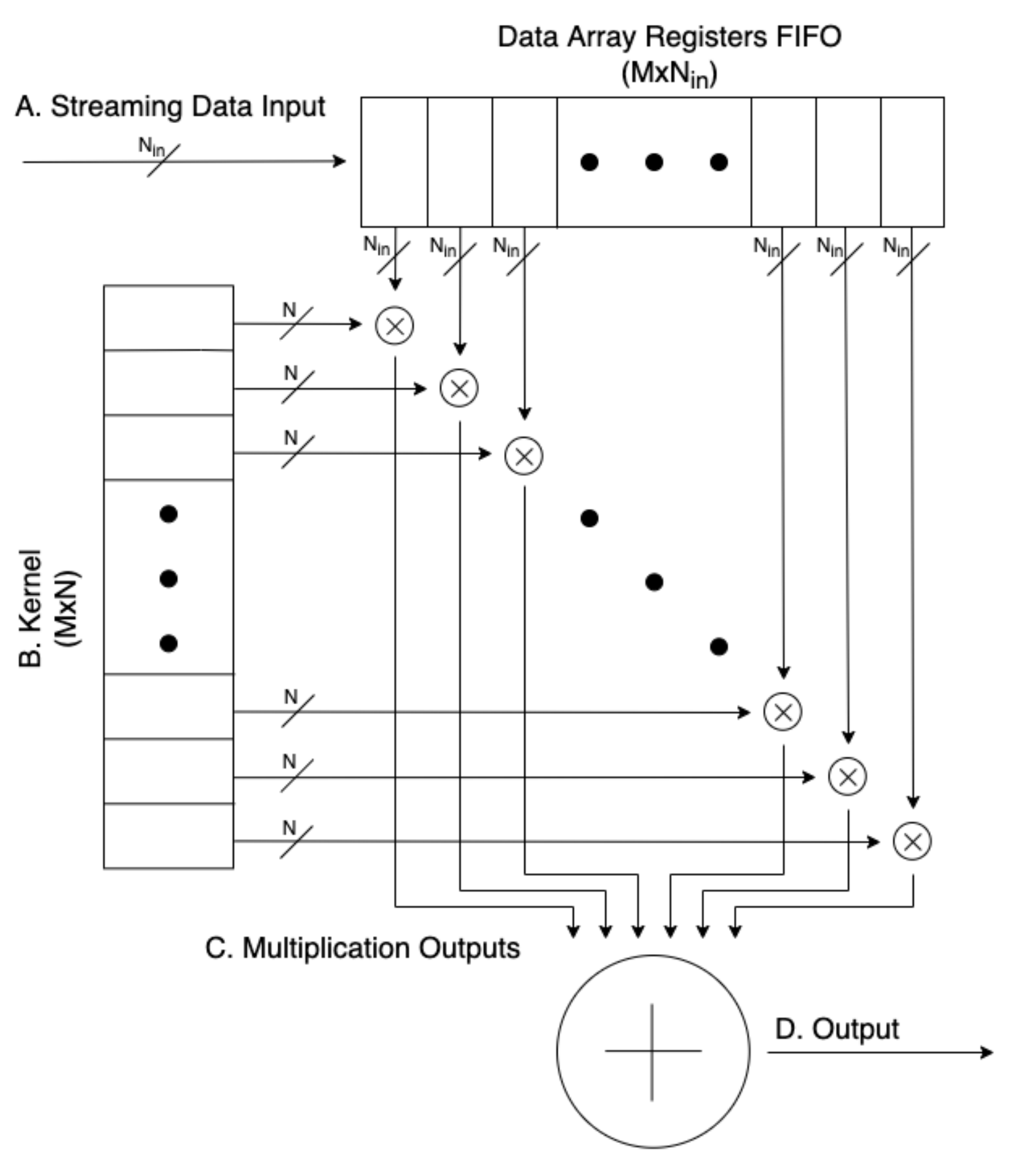}.
    \caption{Data flow for convolution method for derivative. Kernel is stored as an array with $M$ words each of $N$ bits. Each word gets multiplied by the corresponding word in the $M$ word long data array registers which itself is a FIFO with the input data streaming in. Each word of this input data is $N_{in}$ bits. These multiplications are then added together, and the result becomes one element of the derivative output. The points A, B, C, and D correspond to table \ref{tab:conv_wordsize}.}
    \label{fig:convMethod}
    
\end{figure}

Fig. \ref{fig:convMethod} shows how we implement this algorithm.
To perform the convolution on data streaming in, we consider the kernel to be limited in extent to a particular window size, $M$.
Then we load our data into an array of registers equal in extent to the window size.
The data is loaded in first-in first-out fashion (FIFO).
Each element of the array is then multiplied by the corresponding element in the kernel and the result is summed.
The summed result is then one of the output elements of the derivative. 
This is repeated for all the data that is streamed in, with a new element in the total convolution output available after each new piece of streaming data arrives.
This convolution then takes the form of
\begin{equation}
    y_k = \sum_{m=0}^M z_{k,m}*h_m
\end{equation}
where $M$ is the window size of the kernel, $z$ is the array of registers piping the data in and out, and $h$ is the stationary kernel.
In this representation of convolution, for each new $k$, there is a new $z$ since a new piece of data is streamed in. 
Therefore $z$ has a dependence on $m$ and $k$.

The final output will be valid once all of the data has been streamed through the array of registers.
The exact window size and kernel width is application dependent and can be tuned ahead of time based on typical or expected feature widths. 
Once the kernel is chosen, it can be stored in memory as a look-up table (LUT) to be used while the algorithm is running. 

\subsection{Discrete Cosine and Sine Transform Method (DCSTM)}
The streaming Discrete Cosine and Sine Transform Method (DCSTM) is designed to take the derivative of incoming data in a way that produces a low noise output, minimizes resource usage in the FPGA, and has high throughput and low latency. 
One commonly used method for calculating derivatives of functions is to use a fourier transform (FT) where one takes the FT, multiplies by a value proportional to the frequency vector, and then takes the inverse fourier transform (IFT). 
However, fourier transforms can both be costly in an FPGA and be an overkill in terms of what is required \cite{Johnson}. 

Here instead we use the DCT, IDCT, DST, and IDST in a way that resembles an FT \cite{Johnson, buskirk}.
However, for our application we cannot simply feed our data vector in.
Our data is streaming in at a high rate (6 Gs/s) and may consist of over 5120 words. 
This could pose a problem for implementing these aforementioned transforms in an FPGA while retaining high throughput and low latency since it would require a large number of multipliers.
Our algorithm instead handles incoming data in chunks which we describe as a window size. 
Tuning this window size is important and depends on the nature of the expected data.

\begin{figure}[ht!]
    \centering
    \includegraphics[width=0.84\textwidth]{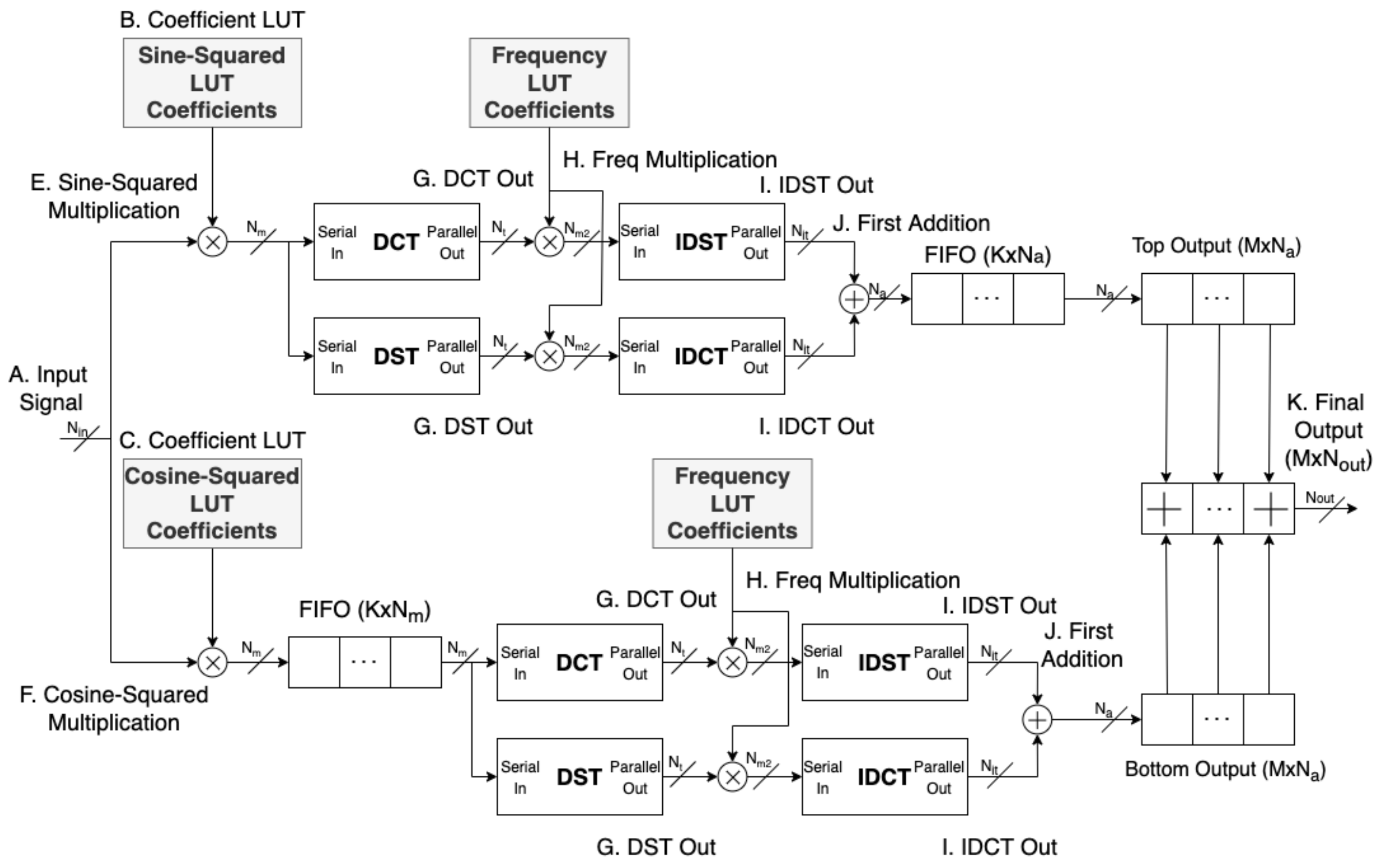}
    \caption{Shows the data flow (left to right) for the derivative algorithm for peak identification. The original signal and a copy are multiplied by SIN$^2$ and COS$^2$ coefficients which are stored in LUTs prior to runtime. Both paths follow identical operations involving DCT and DST, multiplication proportional to frequency, and inverse DST and DCT, and finally summation. One path is delayed at the beginning of the operations and the other path is delayed following the operations in order to regain time coherence for the final, recombining addition. Points A--C and E--K correspond to table \ref{tab:wordsize} while D is in Fig. \ref{fig:DCTAlg}.}
    \label{fig:DerivativeDCTAlg}
\end{figure}

Furthermore, the algorithm must be able to handle data that lands on the boundaries of windows. 
To handle this, we introduce a design that uses two paths for the incoming signal where each path initially applies a different filter, one path then delays its signal, both paths apply identical derivative transformations, the other path then delays its signal, and finally the two paths are recombined. 
Fig. \ref{fig:DerivativeDCTAlg} shows the details of the algorithm.

In the block diagram on the top path, data streams in one word at a time, with each word multiplied by a particular value from a sine-squared (SIN$^2$) filter which was stored pre-runtime in a LUT.
The period of the SIN$^2$ coefficient is proportional to the window size.
The data splits paths again, one into a DCT block and the other to a DST block. 
These blocks take this streaming data and, after a window size amount of data comes in, the output is ready in parallel. 
Once the array of data leaves these blocks, each output word is multiplied by a value proportional to frequency. 
These values then go through the corresponding inverse transform blocks before being added back together. 
The DCT, DST, IDCT, and IDST blocks are described in section \ref{sec:SDCT}.

The other data path that was not multiplied by the SIN$^2$ coefficients is instead multiplied by the cosine-squared (COS$^2$) equivalent. 
Then the data is delayed before going through the same series of transformations. 
This delay is to address capturing signal that might lie on the boundaries of the windows by providing an offset in time to move the signal away from the window size boundary in order to perform the algorithm.
The data on this path then proceeds through the same series of transformations.

Finally the two paths are recombined by first delaying the original path in order to sync the two paths together, regaining temporal coherence. 
They are then added together and to form the final result. 
The multiplication by SIN$^2$ and COS$^2$ values both rolls-off the signal at the edges of the window and ensures the final recombination of the two paths leads to the proper value. 

\subsection{Streaming Discrete Cosine Transform Method (SDCTM)}\label{sec:SDCT}
The normalized DCT, IDCT, DST, and IDST (eq \ref{eq:dct}--\ref{eq:idst}, respectively) are  
\begin{equation}\label{eq:dct}
    \begin{split}
    y_k = 2\sum_{n=0}^{N-1}f x_n \cos \left(\frac{ \pi k(2 n +1)}{2N} \right)\cr
\text{where} \ f =  \begin{cases}
   \sqrt{\frac{1}{4 N}},& \text{if } k = 0\\
     \sqrt{\frac{1}{2 N}},& \text{otherwise } \end{cases}
\end{split}
\end{equation}

\begin{equation}\label{eq:idct}
    \begin{split}
    y_k = f x_0 + 2\sum_{n=1}^{N-1}f x_n \cos \left(\frac{ \pi n(2 k +1)}{2N} \right)\cr
\text{where} \ f =  \begin{cases}
   \sqrt{\frac{1}{ N}},& \text{if } n = 0\\
     \sqrt{\frac{1}{2 N}},& \text{otherwise } \end{cases}
\end{split}
\end{equation}

\begin{equation}\label{eq:dst}
    \begin{split}
    y_k = 2\sum_{n=0}^{N-1}f x_n \sin \left(\frac{ \pi (k+1)(2 n +1)}{2N} \right)\cr
\text{where} \ f =  \begin{cases}
   \sqrt{\frac{1}{4 N}},& \text{if } k = 0\\
     \sqrt{\frac{1}{2 N}},& \text{otherwise } \end{cases}
\end{split}
\end{equation}

\begin{equation}\label{eq:idst}
    \begin{split}
    y_k = f (-1)^k x_{N-1} + &2\sum_{n=0}^{N-2}f x_n \sin \left(\frac{ \pi (2 k +1)(n+1}{2N} \right)\cr
&\text{where} \ f = 
     \sqrt{\frac{1}{2 N}}
\end{split}
\end{equation}

To adapt these equations for streaming data, we first realize each equation as data multiplied by coefficients that depend on the specific $n$ and $k$.
These coefficients then can be packaged into a matrix that is $n\times k$.
By proper multiplication and addition, these functions can all be implemented. 

\begin{figure}[h!]
    \centering
    \includegraphics[width=0.95\textwidth]{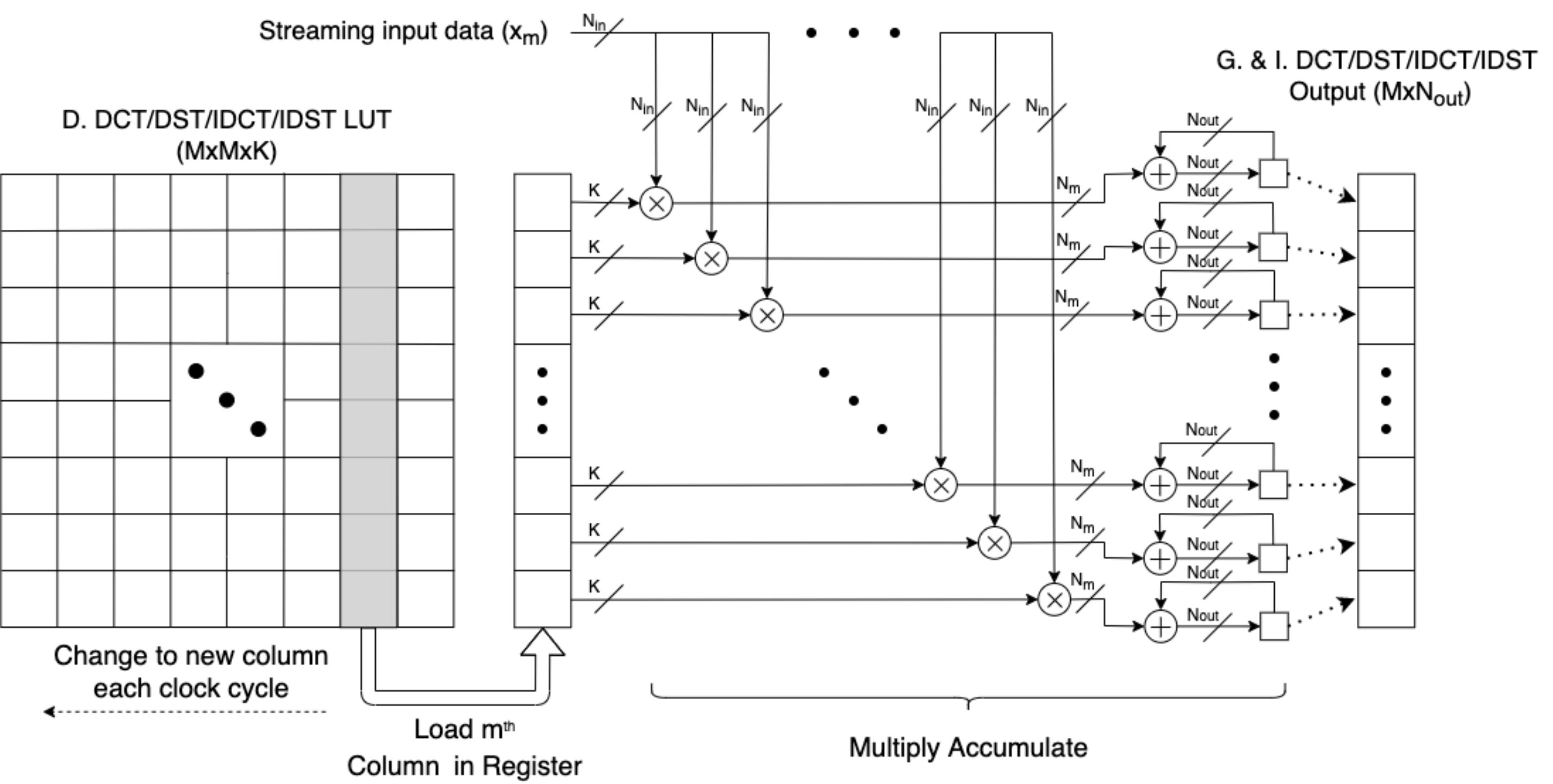}
    \caption{This figure shows the streaming DCT, DST, IDCT, and IDST algorithms which all use this same general framework. Pre-runtime, a $M\times M$ LUT with words of $K$ bits specific to the transform of choice is stored in memory. Each time a new piece of data streams in, a new column from the LUT is loaded into a register array. This data is simultaneously multiplied with each element in this register array. The output of this then goes into a multiply-accumulate function until $M$ pieces of data have streamed in, at which point the process is completed. The points D, G, and I correspond to table \ref{tab:wordsize} and are used within Fig. \ref{fig:DerivativeDCTAlg}.}
    \label{fig:DCTAlg}
\end{figure}

Fig. \ref{fig:DCTAlg} shows the block diagram for these algorithms and the desired transform is selected by choice of LUT matrix for the coefficient values. 
We start before run-time by generating this coefficient matrix according to equations \ref{eq:dct}--\ref{eq:idst} for the corresponding values of $n$ and $k$. 
At the top of the diagram we see the data ($x_{\mbox{in}}$) streaming in one word at a time. 
At each clock cycle, a new word will stream in and the next column from the coefficient matrix will get loaded in parallel into the $M$ long array of registers. 
So each time a piece of data comes in, that specific piece of data will get multiplied $M$ times and accumulate in a temporary array until $M$ pieces of data have come in and the final output vector is ready. 
In total, $M$ numbers of multiplications and additions will occur at each clock cycle (this can be pipelined if needed) as well as loading the next row of the coefficient matrix.
In this way, no clock cycles are wasted since $M$ words come in and the operations are happening in sync with the data streaming in. 

Of course a number of trade-offs are present in this design. 
For reducing resource usage, it is best to limit the size of $K$ as this limits the number of simultaneous multiplications that need to occur (as well as additions). 
Additionally, constraints on the clock period can be loosened if pipelining at various stages is employed, but the ability to do this depends on latency requirements. 
Furthermore, we enforce symmetry and anti-symmetry in the signals streaming into the DCT and DST functions, respectively. 
Doing so takes advantage of the even and odd properties of these functions
In software, this is accomplished by replicating data and flipping it where needed.
In hardware, this is accomplished by assigning 0 to particular computations as either the even or odd coefficient outputs will be 0 depending on if the operation is a DCT or a DST. 
Therefore, those computations can be ignored and will not add to the overall required number of multiplications for the DCT and DST. 

\subsection{Zero Crossing}

\begin{figure}[h!]
    \centering
    \includegraphics[width=0.70\textwidth]{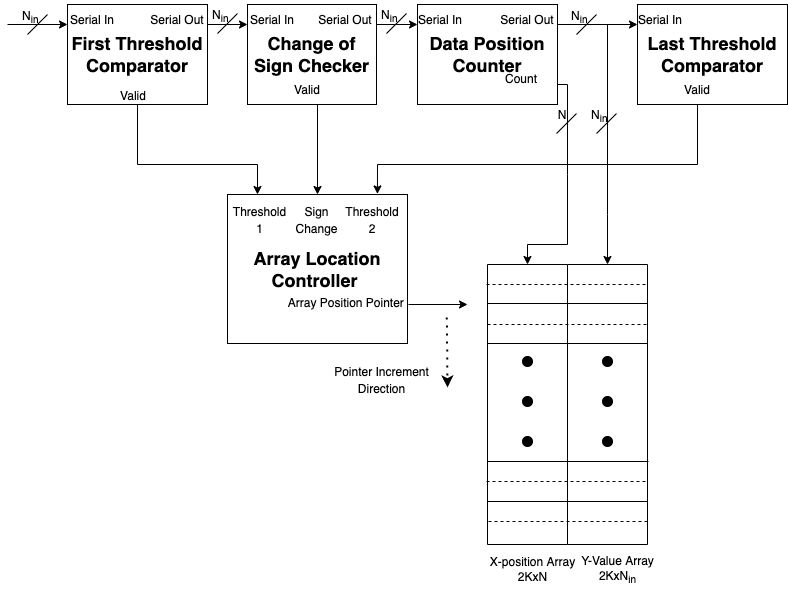}
    \caption{Shows the process for qualifying derivative signals as a valid pulse. Three checks occur on the data. 1) threshold to see that the data goes high (low) enough; 2) checks for a change in sign from positive (negative) to negative (positive); and 3) checks that the data meets an opposite threshold that is low (high) enough. If data meets the change of sign check, the two points around this change of sign are loaded into the position and value arrays. If final disqualifies pulse, these now-invalid values will be overwritten.}
    \label{fig:peakFind}
\end{figure}
The zero crossing algorithm is the final stage of finding the peaks in the data.
Here we search for the x-positions--time for xFEL or frequency for ECE cases--where the derivative signals cross zero.
One possible resource conserving approach to this search for streaming data is shown in Fig. \ref{fig:peakFind}.
Given that features resemble something between a triangle and a Gaussian pulse, the derivative will also have a known shape. 
With this expectation, the algorithm waits for the streaming derivative input signal to pass a particular threshold, then change sign, and finally cross a second threshold.
If the derivative meets these criteria, then the two positions in the data that were closest to the zero crossing are saved.
In hardware, this would involve an array of predefined length for holding all of the data pairs and corresponding positions for qualifying zero crossings. 
A pointer in the array would then update where in the array the next data pair will be saved. 
In this way, if a data pair meet the first two qualifications threshold and then fail the final threshold, the saved data pair from this can be overwritten by moving the pointer back.
This method determines two x-axis values and two corresponding y-axis values for the points surrounding the zero crossing. 

The last portion of the algorithm arrives at an estimate for the zero crossing based on these two surrounding points.
In the simplest approach, the position corresponding to the value with smaller magnitude between the two can be taken as the x-axis position stamp for the zero crossing.
We call this method the direct method for finding the zero crossing.

\begin{algorithm}[h!]
\caption{Weighted average zero crossing calculation. ($t1$,$v1$) is the first point and ($t2$,$v2$) is the second point.}\label{alg:weighted_avg}
\begin{multicols}{2}
\begin{algorithmic}[1]
\Require $(v1 \leq 0 \land v2 \geq 0) \lor (v2 \leq 0\land v1 \geq 0)$
\If{$|v1| == 0$}
    \State $t_{out} \gets t1 $ 
\ElsIf{$|v2| == 0$}
    \State $t_{out} \gets t2 $
\Else
    \If{$|v1| \geq |v2|$}
        \If{$|v1|\leq 2*|v2|$}
            \State $t_{out} \gets (2*t2+t1)/3 $ 
        \ElsIf {$|v1|\leq 4*|v2|$}
            \State $t_{out} \gets (4*t2+t1)/5 $ 
        \ElsIf {$|v1|\leq 8*|v2|$}:
            \State $t_{out} \gets (8*t2+t1)/9 $ 
        \ElsIf {$|v1|\leq 16*|v2|$}:
            \State $t_{out} \gets (16*t2+t1)/17 $ 
        \Else:
            \State $t_{out} \gets t2 $ 
        \EndIf
    \ElsIf{$|v1| == |v2|$}
        \State $t_{out} \gets (t1+t2)/2 $ 
    
    \Else
        \If{$|v2|\leq 2*|v1|$}
            \State $t_{out} \gets (2*t1+t2)/3 $ 
        \ElsIf {$|v2|\leq 4*|v1|$}
            \State $t_{out} \gets (4*t1+t2)/5 $ 
        \ElsIf {$|v2|\leq 8*|v1|$}:
            \State $t_{out} \gets (8*t1+t2)/9 $ 
        \ElsIf {$|v2|\leq 16*|v1|$}:
            \State $t_{out} \gets (16*t1+t2)/17 $ 
        \Else:
            \State $t_{out} \gets t1 $ 
        \EndIf
    \EndIf
\EndIf
\end{algorithmic}
\end{multicols}
\end{algorithm}

A second method uses a weighted average between the two surrounding points to elicit a better resolved x-position stamp. 
However, a standard weighted average is not necessarily the most conducive option for efficient implementation in an FPGA. 
Instead we propose a modified weighted average method.
This method, shown in algorithm \ref{alg:weighted_avg}, starts by assuming that by construction one point has negative value ($v1$) and the other has positive value ($v2$).
Then it checks if either the $y$ value of the first point (called $v1$) or the $y$ value of the second point (called $v2$) are zero.
If not, then it compares $|v1|$ and $|v2|$ in several steps. 
If $|v1|$ is larger than $|v2|$, then it checks how much larger.
Based on how much larger, the output position ($t_{\mbox{out}}$) is assigned a corresponding proportion of $t1$ and $t2$.
Specifically, we check if $|v1|$ is 2 times $|v2|$, 4 times $|v2|$, 8 times $|v2|$, or 16 times $|v2|$.
This results in simple, digital operations in the FPGA design.
Similarly, we do the same comparisons for the case where $v2$ is bigger than $v1$, and we handle the case where $v1$ and $v2$ are the same value by a standard average. 
Additionally, the output position estimates can then be adjusted based on known adjusting constants for time-zero or other required shifts. 
This method leverages the ability to easily compare two values and the corresponding factors involving powers of two, which in an FPGA only requires shift-registers and comparators. 
The denominators for division can then be stored in reciprocal form in LUTs ahead of time for multiplication for the final output.

%% file: sections/results.tex
In this section, we present the simulation results for the convolution method and the DCSTM for both the derivative and the final peak location estimates. 
We start with individual subsections on convolution and DCSTM showing how we selected algorithm specific parameters. 
Then we show the comparison in performance for the two algorithms on two test data use-cases-- the Cookiebox time-of-flight data and the fusion ECE data. 
For the peak location comparisons, we use fabricated data made from Gaussian peaks placed at specific locations so that we can precisely analyze our algorithms. 
We compare both the direct method and the weighted average method for the final peak location estimates. 

\subsection{Convolution Method}
\label{sec:convResults}
The convolution method requires the kernel to be chosen ahead of run time. 
Specifically, the kernel length and the width need to be selected based on expected data.
The kernel width, in particular, must be able to capture the width of the typical pulses. 
Furthermore, we want our kernel to be continuous and continuous differentiable. 
To find a function of this nature, we can start by considering a kernel that mimics the derivative functionality, such as a step function. 
Then we can inspect the noiseless power spectrum of our data. 
We can Wiener filter the data to get an approximation to the noiseless power spectrum and then multiply this by the Fourier transformed step function. 
Taking the inverse Fourier transform results in pulses that closely resemble the Wiener filtering derivative of a Gaussian pulse. 
We can then choose the width of the kernel, which we had originally chosen to be this derivative of the Gaussian pulse, to match the inverse of the Wiener filtered power spectrum. 


\begin{figure}[htb!]
    \centering
    \begin{subfigure}{1\textwidth}
		\includegraphics[width=0.9\textwidth]{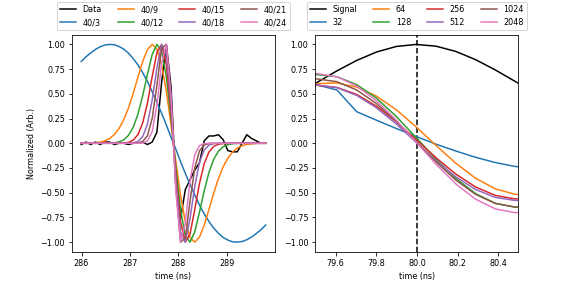}
	\end{subfigure}
	
	\begin{subfigure}{.3\textwidth}
		
		\caption{}
		\label{fig:kernelMatch}
	\end{subfigure}
	\begin{subfigure}{.43\textwidth}
		
		\caption{}
		\label{fig:windowsize}
	\end{subfigure}
    \caption{Figure shows pre-runtime tuning operations for both the convolution method (a) and DCSTM (b). Specifically (a) compares the the inverse transformed Wiener filtered power spectrum multiplication output (in black) and convolution kernels with window size of 40 and widths ranging from 40/3 to 40/24. (b) shows a Gaussian pulse (labeled Signal in black), a dashed line indicating the peak position of the signal, and seven versions of the derivative of the signal all using the DCSTM algorithm but with different window sizes.}
    \label{fig:convDcstm}
\end{figure}


Fig. \ref{fig:kernelMatch} shows derivative Gaussian kernels with a window size of 40 with Gaussian widths varying from 40/3 to 40/24. 
These are overlaid on the output from the inverse transformed Wiener filtered power spectrum multiplication output (labeled as Data in black).
We can see that the kernel and the output overlap closely as the kernel width gets close to 40/18, and this kernel would overlap even better if the signal were truly noise free.  
This suggests that such a filter choice would select out the derivative of the data without emphasizing noise, and one can tune the width of the kernel to match the expected data appropriately.


For running the convolution derivative method, apart from the kernel width, the word sizes at each point in the algorithm must be chosen for the fixed point representation. 
Table \ref{tab:conv_wordsize} shows the selected word sizes corresponding to the different points in the convolution block diagram (see Fig. \ref{fig:convMethod}).
The final output word size is 25 bits with 10 fractional bits for the Cookiebox data and just 25 integer bits for the fusion ECE data. 
The particular word sizes chosen resulted in acceptable errors for the given use-cases.
\begin{table}[h!]
    
\centering
\begin{tabular}{|c|c | c |c||c| c|} 
 \hline
 && \multicolumn{2}{|c||}{Cookiebox (bits)}&\multicolumn{2}{|c|}{Fusion (bits)} \\ [0.5ex] 
 \hline
 Label&Description &Total& Fractional&Total& Fractional \\
 \hline
 (A)& Input Signal & 12 & 7 &13&0 \\ 
 \hline
 (B)&Kernel & 11 & 2 &10&2 \\ 
 \hline
 (C)&Multiplication Outputs & 24 & 10&22&0 \\ 
 \hline
 (D)&Final Output &25&10&25&0 \\
  \hline 

\end{tabular}

\caption{Table shows the word sizes at each stage in the Convolution algorithm for both Cookiebox and fusion data. For each word, the total bits and the number of bits after the decimal (fractional bits) for the fixed point representation are shown. The labels for these words can be matched to the block diagram of the algorithm corresponding to the letter designator (see Fig. \ref{fig:convMethod}).}
\label{tab:conv_wordsize}
\end{table}

\subsection{DCSTM}

The DCSTM algorithm requires the window size to be chosen ahead of run time.
Additionally, the word width size--the number of bits in each word--for each stage of the algorithm must be chosen. 
All of these parameters can be tuned and are dependent on the expected data.  
The window size selection depends on the nature of the typical data, mostly the expected pulse widths. 
For our data we looked at how window size would affect location of the zero crossing for the derivative.
Fig. \ref{fig:windowsize} shows how the window size can affect the derivative. This figure shows the original signal (black), a dashed black line indicating the peak location of the signal, and the derivatives calculated using DCSTM for window sizes from 32 to 2048 words for test data of similar width to our real data.


The window size needs to be larger if the pulse is wide because we must be able to capture significant portions of the pulse. 
We can see that for 128 and beyond, the derivative signals begin to converge. 
Therefore, we select 128 as the window size as we want as small of a window size as possible while still retaining accurate enough zero crossing information. 

After having selected window size, bit width of the data words for each section of the algorithm were determined.
We tested our selected values with test data sets from both use-cases considered in this paper.

\begin{table}[h!]
    
\centering
\begin{tabular}{|c|c | c |c||c| c|} 
 \hline
 && \multicolumn{2}{|c||}{Cookiebox (bits)}&\multicolumn{2}{|c|}{Fusion (bits)} \\ [0.5ex] 
 \hline
 Label&Description &Total& Fractional&Total& Fractional \\
 \hline
 (A)& Input Signal & 12 & 7 &13&0 \\ 
 \hline
 (B)&Sine-Squared LUT & 9 & 8&12&10 \\ 
 \hline
 (C)&Cosine-Squared LUT & 9 & 8&12&10 \\ 
 \hline
 (D)&DCT/DST/IDCT/IDSCT LUT & 20&18&12&8 \\
 \hline
 (E)&Sine-squared Multiplication & 15&10 &20&5\\
 \hline
 (F)&Coine-squared Multiplication & 15&10 &20&5\\
 \hline
 
 (G)&DCT/DST Outputs & 20&8&20&0 \\
 \hline
 (H)&Frequency Multiplication Outputs & 28&8 &22&0\\
 \hline
 (I)&IDCT/IDST Outputs & 25&10 &23&0\\
 \hline
 (J)&First Set Additions & 26&10&24&0\\
 \hline
 (K)&Final Output &27&10&25&0\\
  \hline

\end{tabular}
\caption{Table shows the word sizes at each stage in the DCSTM algorithm for both Cookiebox and fusion ECE data sets. For each word, the total bits and the number of bits after the decimal (fractional bits) for the fixed point representation are shown. The labels for these words can be matched to the block diagram of the algorithm corresponding to the letter (see Fig. \ref{fig:DerivativeDCTAlg} and \ref{fig:DCTAlg}).}
\label{tab:wordsize}
\end{table}

Table \ref{tab:wordsize} shows the chosen values for the word sizes, which correspond to the associated points in the algorithm block diagram, Fig. \ref{fig:DerivativeDCTAlg} and \ref{fig:DCTAlg}.
For the Cookiebox data, the final output word is 27 bits in total with 10 fractional bits. 
For the fusion data, the final output word is 25 integer bits. 
These word sizes give errors on the derivative calculation that are acceptable for these applications.

\subsection{Convolution and DCSTM Comparisons}
\begin{figure}[hb!]
    \centering
    \begin{subfigure}{1\textwidth}
		\includegraphics[width=0.9\textwidth]{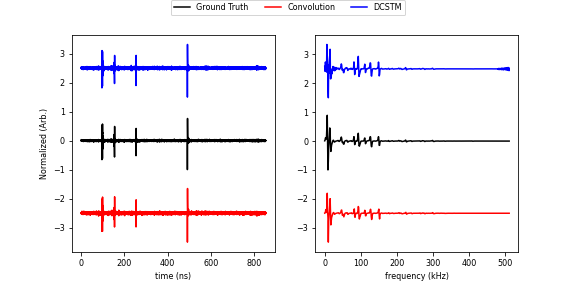}
	\end{subfigure}
	\vspace{-20pt}
	\begin{subfigure}{.3\textwidth}
		
		\caption{}
		\label{fig:full_deriv_cookie}
	\end{subfigure}
	\begin{subfigure}{.45\textwidth}
		
		\caption{}
		\label{fig:full_deriv_fusion}
	\end{subfigure}
	
	\begin{subfigure}{1\textwidth}
		\includegraphics[width=0.9\textwidth]{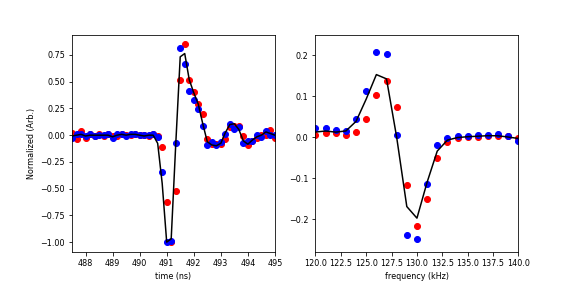}
	\end{subfigure}
	\begin{subfigure}{.3\textwidth}
		
		\caption{}
		\label{fig:full_deriv_zoom_cookie}
	\end{subfigure}
	\begin{subfigure}{.45\textwidth}
		
		\caption{}
		\label{fig:full_deriv_zoom_fusion}
	\end{subfigure}
		
		

    \caption{Figure shows the derivative of the data collected from the Cookiebox detector-- (a) and (c)-- and from the fusion data-- (b) and (d). The ground truth (black) is the derivative calculated using a standard software package. The remaining signals are the derivatives calculated using the DCSTM algorithm (blue) and using the convolution method (red). For visibility, (a) and (b) have the DCSTM and convolution method offset. (c) and (d) show the results zoomed on one pulse and with discrete points for the derivatives to showcase how the derivative outputs affect the zero crossing algorithms.}
    \label{fig:combo_plot}
\end{figure}

For both algorithms, the pre-runtime parameters have to be set based on the expected characteristics of the data. 
First we inspect the derivative functionality of both algorithms.
Fig. \ref{fig:full_deriv_cookie} shows the convolution method derivative and DCSTM derivative applied to the Cookiebox dataset, and Fig. \ref{fig:full_deriv_fusion} shows the two methods applied to the fusion spectral data.
Here the ground truth derivative (black) is in the center of the figures and the convolution method (red) is negatively offset while the DCSTM (blue) is positively offset for better visibility.
Visually, the two algorithms align well with the ground truth derivative.

To quantify the performance of the algorithms we take the mean-squared error with the ground truth.
For the Cookiebox data, the DCSTM algorithm has superior performance with an error of 1.23 while the convolution method has an error of 3.81.
For the fusion data, the DCSTM also has superior performance with an error of 1.03 while the convolution method has an error of 1.26.

For further inspection, we can focus around one of the peaks in the data.
Fig. \ref{fig:full_deriv_zoom_cookie}  and \ref{fig:full_deriv_zoom_fusion}--CookieBox and fusion ECE, respectively--show the ground truth (black) and discrete points of the convolution-based derivative (red) and DCSTM derivative (blue).
In both Fig. \ref{fig:full_deriv_zoom_cookie} and \ref{fig:full_deriv_zoom_fusion}, we see that there is a slight horizontal offset of the convolution derivative output that does not show up in the DCSTM algorithm. 
This offset affects the points closest to the zero crossing and therefore will have an effect on the estimation methods for the peak-finding algorithms.

		
		
    

\begin{figure}[htb!]
    \centering
    \begin{subfigure}{1\textwidth}
		\includegraphics[width=0.9\textwidth]{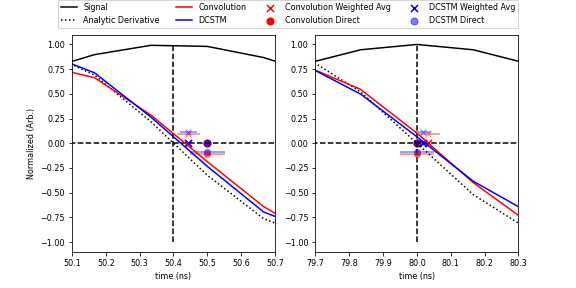}
	\end{subfigure}
	
	\begin{subfigure}{.3\textwidth}
		
		\caption{}
		\label{fig:fab1}
	\end{subfigure}
	\begin{subfigure}{.45\textwidth}
		
		\caption{}
		\label{fig:fab2}
	\end{subfigure}
    \caption{Figure shows two pulses out of the total generated dataset. The signal is shown in black with a dashed line indicating the peak location. The true derivative is the analytic derivative of the Gaussian shown as a dotted black line. The convolution derivative is shown in red, and the DCSTM derivative is shown in blue. The corresponding peak estimates are shown as ``x" for the weighted average and a ``o" for the direct method. All of the signals have been normalized based on the maximum of the absolute value of the respective signal (across all pulses in that signal). (a) shows a pulse that does not have its center on the time grid. (b) shows a pulse that does lie perfectly on the time grid. For each zero crossing estimate, error bars based on the root mean-squared error are included. The error bars are offset from their associated points for better visibility. The RMSE for convolution weighted average is 33.8 ps, for convolution direct is 51.4 ps, for DCSTM weighted average is 24.4 ps, and for DCSTM direct is 51.4 ps.}
    \label{fig:zero_crossing}
\end{figure}

To better analyze how the offset and the discretization affect the zero crossing estimate, we use a generated dataset to test the peak-finding algorithms.
For this we use Gaussian pulses placed at specific locations.
We can also get the analytic derivative of the Gaussian pulses for comparisons with the algorithms. 
Using both the convolution method and the DCSTM, we can test both zero crossing methods.
Fig. \ref{fig:zero_crossing} shows two pulses in the dataset generated using these Gaussian pulses.
Here the signal is shown in black, the analytic derivative of the Gaussian is shown as a dotted black line, the convolution method derivative is in red, the two estimates for the zero crossing--direct and weighted average--based on the convolution method are shown in red, the DCSTM derivative is in blue, and the two estimates for the zero crossing algorithm based on the DCSTM are shown in blue.
The first pulse (Fig. \ref{fig:fab1}) shows an example where the peak is located off the exact time grid.
The second pulse (Fig. \ref{fig:fab2}) shows an example where the peak is located on the time grid. 

In both Fig. \ref{fig:fab1} and \ref{fig:fab2}, we see that the convolution derivative method has a larger shift from the analytic derivative. 
In both cases, the weighted average zero crossing has a better estimate for the zero crossing of that derivative, and this means that the peak estimate based on this method will depend how closely aligned that derivative is with the analytic derivative. 
For Fig. \ref{fig:fab1} where the peak is not on the time grid, the direct method zero crossing estimate is very poor, yielding a result nearly 0.1 ns wrong for both the convolution and DCSTM. 
However, in Fig. \ref{fig:fab2} where the peak is perfectly on the time grid, the direct method yields a very good result.
This discrepancy means that the direct method is not guaranteed to deliver a valid result while the weighted average method will yield a reasonable result even for peaks that are not exactly on the grid.

To better quantify these differences, we use root mean-squared error (RMSE) for the estimated peak locations compared to their actual locations for the two versions of the peak finding for both derivative methods as given by
\begin{equation}
    \label{eq:rmse}
    RMSE = \sqrt{\frac{1}{N}\sum^{N-1}_{i=0} (x_i - x_{\mbox{truth}})^2}
\end{equation}
where $N$ is the number of peaks identified in the data and $x_{\mbox{truth}}$ is the true peak location.
For this generated data, some of the peaks are located on the grid and others are not.
The resulting errors are shown as error bars in Fig. \ref{fig:zero_crossing} for the identified zero crossings. 
We see that the errors for the direct methods for both DCSTM and convolution methods are nearly identical and are nearly twice as large as the errors for the weighted average methods for both DCSTM and convolution.
This means that even in the cases where the direct method happens to yield a closer estimate to the zero crossing because of the peak location being on a grid point, the associated error bar still gives a relatively high uncertainty.
The DCSTM method using the weighted average zero crossing estimate gives the lowest RMSE of 24.4 ps for generated data in this dataset. 
This high resolution is in fact a factor of 2 better than that required for 0.25~eV spectral resolution with 500~eV electrons as indicated in Fig.~7 of Ref.~\cite{walter2021multi}.

%% file: sections/conclusions.tex


Systems requiring live decision making, fast feedback and control, and machine learning-optimized data reduction all rely on data being fed in and pre-processed quickly and accurately.
One such application is peak finding for dimensional reductive featurization of raw data.
This particular application has been required in a wide range of experiments in topics spanning nuclear physics to biological applications, and the approach has involved both analog electronics with constant fraction discriminators and iterative curve fitting algorithms in software.
Leveraging the computational efficiency of older approaches is seeing a resurgence given modern needs for high throughput and low latency data processing.
This encourages the use of hardware-based algorithm implementations, such as on an FPGA, where highly parallel computations and real-time processing can bring the best of traditional analog approaches to more generalized digital applications in streaming data featurization.
Furthermore, hardware-based approaches enable pipelined integration with detector and data acquisition systems and can be placed intimately close to the detectors and digitizers. 
This is the frontier of ML at-the-edge that drives in-device edge computing initiatives.

In this paper we have presented and compared two featurization algorithms designed for implementation in hardware but simulated in software. 
For both of these algorithms we discuss the fundamental design, the parameter selection pre-runtime, and the final results for both generated data and for real data from our use cases.
We focus on two distinct use cases for in-flight streaming data processing: a) X-ray pulse reconstruction at SLAC’s LCLS-II X-ray Free-Electron Laser and b) ECE as a plasma diagnostic at the DIII-D tokamak fusion reactor. 
The results from both show that we are able to calculate a derivative and find the zero crossings in that signal (therefore the centroids of the associated peaks) using both a convolution method and a discrete cosine and sine transform method all while keeping our bit representations quite low. 

The convolution method requires, for our use cases, 40 simultaneous multiplications followed by an addition of all of those outputs. 
However, we are able to keep the final output size to roughly only double the initial input word size of the data.
Moreover, the computations are all performed without looping or bottlenecks such that the data can continuously feed the streaming pipeline. 
For the DCSTM, the method requires, for our use cases, between roughly 250 and 1000 simultaneous multiplications. 
To keep full precision, the output bitsize, in theory, would need to be very large.
Furthermore, a native version of the DCSTM that uses standard discrete sine and cosine transforms would not be able to operate on streaming data.
However, our novel method of windowing the signal and using smaller discrete sine and cosine transforms to compute the total allows us to operate on streaming data while reducing the number of multipliers from roughly 5000 down to 1000 for the CookieBox data, for example.
Additionally, our output word size is also roughly double the input word size even with all of these multiplications, so the hardware resources are still manageable.

The DCSTM algorithm gives a better result than the convolution method. 
However, the DCSTM requires significantly more resources in an FPGA.
This trade-off means that applications requiring higher accuracy may choose the DCSTM but will need larger hardware to accomplish this task. 
We are currently working on the FPGA implementation of the DCSTM algorithm, and future work will contain a comparison, in hardware, of these two algorithms. 
In addition, next developments will demonstrate combinations of serial streaming and parallel burst versions of these algorithms.
For example, our CookieBox data is sampled at 6 Gs/s while our FPGA clock runs slower than this sample rate.
Thus, we would have access to batches of data at a given time.
We are beginning to work on versions of our algorithms that will leverage this parallel data availability and vectorize some of the operations to perform streaming-vectorized operations for further reduction in latency. 

%% file: main.bbl
\begin{thebibliography}{6}

 
 \bibitem {cookiebox}
  Therrien, A.C., Herbst, R., Quijano, O., Gatton, A., Coffee, R.: Machine learning at
the edge for ultra high rate detectors. IEEE Nuclear Science Symposium and Medical
Imaging Conference, 10, pp. 1–4 (2019).


\bibitem {toka}
Lazarus, E.A., Chu, M.S., Ferron, J.R., Helton, F.J., Hogan, J.T., Kellman, A.G., Lao, L.L., Lister, J.B., Osborne, T., Snider, R. and Strait, E.J.: Higher beta at higher elongation in the DIII‐D tokamak. Physics of Fluids B: Plasma Physics (1991).

\bibitem {peakFind:thesis}
Naaranoja, Tiina: Digital Signal Processing for Particle Detectors in Front-End Electronics. University of Helsinki, Master's Thesis in Physics (2014).

\bibitem {peakFind:old}
Cox, S. A., P. R. Hanley: A Fast Zero-Crossing and Constant Fraction Timing Discriminator with Emitter Coupled Integrated Circuits. IEEE Transactions on Nuclear Science 18.3 (1971).

\bibitem {peakFind:simple}
Wall, R. W.: Simple methods for detecting zero crossing. IECON'03. 29th Annual Conference of the IEEE Industrial Electronics Society. Vol. 3. IEEE (2003).

\bibitem {iterative}
Bevington, P.R., Robinson, K.D.: Data reduction and error analysis. McGraw Hill, New York (2003).
\bibitem{iterPaper}
Brown, Adrian J: Spectral curve fitting for automatic hyperspectral data analysis. IEEE Transactions on Geoscience and Remote Sensing 44.6 (2006).

\bibitem{iterPaperGood}
De Weijer, A. P., Lucasius, C. B., Buydens, L., Kateman, G., Heuvel, H. M., Mannee, H.: Curve fitting using natural computation. Analytical Chemistry 66.1 (1994).

\bibitem{egg}
Maatta, K., Kostamovaara, J.: A high-precision time-to-digital converter for pulsed time-of-flight laser radar applications. IEEE Transactions on Instrumentation and Measurement (1998).

\bibitem{wiener}
Benesty, J., Chen, J., Huang, Y. A., Doclo, S.: Study of the Wiener filter for noise reduction. In Speech enhancement. Springer, Berlin, Heidelberg pp. 9-41 (2005).

\bibitem{Heckbert}
Heckbert, Paul S.: Filtering by repeated integration. ACM SIGGRAPH Computer Graphics 20.4. pp 315-321 (1986).

\bibitem{boxlets}
Simard, P., Bottou, L., Haffner, P., LeCun, Y.: Boxlets: a fast convolution algorithm for signal processing and neural networks. Advances in neural information processing systems, 11 (1998).

\bibitem{Talbi}
Talbi, F., Alim, F., Seddiki, S., Mezzah, I., Hachemi, B.: Separable convolution gaussian smoothing filters on a xilinx FPGA platform. In Fifth International Conference on the Innovative Computing Technology. pp. 112-117 (2015).

\bibitem{website}
Fisher, R., Perkins, S., Walker, A., Wolfart, E.: Zero crossing detector. The University of Edinburgh.

\bibitem{book1}
Gonzalez, R. C., Woods, R. E.: Digital image processing second edition. Beijing: Publishing House of Electronics Industry (2002).

\bibitem{Johnson}
Johnson, S. G., Frigo, M.: A modified split-radix FFT with fewer arithmetic operations. IEEE Transactions on Signal Processing, 55(1), pp. 111-119 (2006).

\bibitem{buskirk}
Lundy, T., Van Buskirk, J.: A new matrix approach to real FFTs and convolutions of length 2 k. Computing, pp. 23-45 (2007).

\bibitem{walter2021multi}
Walter, P., Kamalov, A., Gatton, A., Driver, T., Bhogadi, D., Castagna, J.C., Cheng, X., Shi, H., Obaid, R., Cryan, J. and Helml, W.: Multi-resolution electron spectrometer array for future free-electron laser experiments. Journal of Synchrotron Radiation 28.5 (2021)



 \end{thebibliography}
